\documentclass[preprint,showpacs,preprintnumbers,amsmath,amssymb]{revtex4}
\usepackage{amsmath,amsfonts,amssymb}
\usepackage{epsfig}
\def\bit{\begin{itemize}}
\def\eit{\end{itemize}}
\def\ben{\begin{enumerate}}
\def\een{\end{enumerate}}
\def\bed{\begin{description}}
\def\eed{\end{description}}
\def\b{\beta}

\def\k{\kappa}
\def\l{\lambda}

\def\half{\frac{1}{2}\,}
\def\third{\frac{1}{3}\,}

\def\lsim{\raise0.3ex\hbox{$<$\kern-0.75em\raise-1.1ex\hbox{$\sim$}}}
\def\gsim{\raise0.3ex\hbox{$>$\kern-0.75em\raise-1.1ex\hbox{$\sim$}}}

\let\jnfont=\rm
\def\NPB#1,{{\jnfont Nucl.\ Phys.\ B }{\bf #1},}
\def\PLB#1,{{\jnfont Phys.\ Lett.\ B }{\bf #1},}
\def\EPJC#1,{{\jnfont Eur.\ Phys.\ Jour.\ C }{\bf #1},}
\def\PRD#1,{{\jnfont Phys.\ Rev.\ D }{\bf #1},}
\def\PRL#1,{{\jnfont Phys.\ Rev.\ Lett.\ }{\bf #1},}
\def\MPLA#1,{{\jnfont Mod.\ Phys.\ Lett.\ A }{\bf #1},}
\def\JPG#1,{{\jnfont J.\ Phys.\ G}{\bf #1},}
\def\CTP#1,{{\jnfont Commun.\ Theor.\ Phys.\ }{\bf #1},}
\def\JHEP#1,{{\jnfont JHEP \ }{\bf #1},}
\def\NPPS#1,{{\jnfont Nucl.\ Phys.\ Proc.\ Suppl.\ }{\bf #1},}

\def\beq{\begin{equation}}
\def\eeq{\end{equation}}
\def\bea{\begin{eqnarray}}
\def\eea{\end{eqnarray}}
\newcommand{\ba}{\begin{array}}
\newcommand{\ea}{\end{array}}
\def\nn{\nonumber}

\def\endash{\hbox{--}}
\def\Det{\mathop{\rm Det}}

\begin{document}
\title{A comparative study of dark matter in the MSSM and its singlet extensions:
a mini review}

\author{Wenyu Wang}

\affiliation{
 Institute of Theoretical Physics, College of Applied Science,
 Beijing University of Technology, Beijing 100124, China \vspace{0.5cm}}

\begin{abstract}
In this note we briefly review the recent studies
of dark matter in the MSSM and its singlet extensions:
the NMSSM, the nMSSM, and the general singlet extension.
Under the new detection results of CDMS II, XENON, CoGeNT and PAMELA,
we find that (i) the latest detection results can exclude a large
part of the parameter space which  allowed by current collider constraints
 in these models. The future SuperCDMS and XENON can cover most of the allowed
parameter space; (ii) the singlet sector will decouple
from the MSSM-like sector in the NMSSM,
however, singlet sector makes the nMSSM quite different from the MSSM;
(iii) the NMSSM can allow light dark matter at several GeV exists.
Light CP-even or CP-odd Higgs boson must be present so as
to satisfy the measured dark matter relic density.
In case of the presence of a light CP-even Higgs boson,
the light neutralino dark matter can  explain the CoGeNT
and DAMA/LIBRA results; (iv) the general singlet extension of the
MSSM gives a perfect explanation for both  the relic density and
the PAMELA result through the Sommerfeld-enhanced annihilation.
Higgs decays in different scenario are also studied.
\end{abstract}

\maketitle
\section{Introduction}
Although there are many theoretical or aesthetical arguments for
the necessity of TeV-scale new physics, the most convincing
evidence is from the WMAP (Wilkinson Microwave Anisotropy Probe)
observation of the cosmic cold dark matter, which naturally
indicates the existence of WIMPs (Weakly Interacting Massive
Particle) beyond the prediction of the Standard Model (SM).  By
contrast, the neutrino oscillations may rather imply trivial new
physics (plainly adding right-handed neutrinos to the SM) or new
physics at some very high see-saw scale unaccessible to any
foreseeable colliders. Therefore, the TeV-scale new physics to be
unraveled at the Large Hadron Collider (LHC) is the most likely
related to the WIMP dark matter.

If WIMP dark matter is chosen by nature, it will give a strong
argument for low-energy supersymmetry (SUSY) with R-parity which can
give a good candidate. Nevertheless,  SUSY is motivated
for solving the hierarchy problem elegantly. It can also solve
other puzzles of the SM,  such as the $3\sigma$ deviation of the
muon anomalous magnetic  moment from the SM prediction.
In the framework of SUSY, the most intensively studied model is
the minimal supersymmetric standard
model (MSSM) \cite{mssm}, which is the most economical realization
of SUSY. However, this model suffers from the $\mu$-problem.
The $\mu$-parameter is the only dimensional parameter in the
SUSY conserving sector.  From a top down view, one would
expect the $\mu$ to be either zero or at the Planck scale.
But in the MSSM, the relation of the electro-weak (EW)
scale soft parameters ($\tilde{m}_{d}^2,~\tilde{m}_{u}^2$) \cite{sugrawgc}
\begin{equation}
\half M_Z^2={\tilde{m}_{d}^2 - \tilde{m}_{u}^2 \tan^2\beta \over \tan^2\beta -1} - \mu^2,
\end{equation}
makes that $\mu$ must be at the EW scale, while LEP constraints on the
chargino mass require $\mu$ to be non-zero \cite{lepsusy}.
A simple solution is to promote $\mu$
to a dynamical field in extensions of the MSSM that contain an additional singlet
superfield $\hat{S}$ which does not interact with the MSSM fields other than
the two Higgs doublets.  An effective $\mu$ can be reasonably got
at EW scale when $\hat S$ denotes the vacuum
expectation value (VEV) of the singlet field. Among these extension models
 the next-to-minimal supersymmetric model (NMSSM)
\cite{NMSSM} and the nearly minimal supersymmetric model (nMSSM)
\cite{xnMSSM1,xnMSSM2} caused much attention recently.
Note that the little hierarchy problem which is also a trouble
of the MSSM is relaxed greatly in the NMSSM.

If introduce a singlet superfield to the MSSM, the Higgs sector will have
one more CP even component and one more CP odd component, and the
neutralino sector will have one more singlino component.
These singlet multiplets compose a ``singlet sector'' of the MSSM.
It can make the phenomenologies of SUSY dark matter and Higgs
different from the MSSM. More and more precision results
of dark matter detection
give us an opportunity to test if this singlet sector really exists.
For example, experiments for the underground direct detection
of cold dark matter $\tilde{\chi}$ have recently made significant
progress. While the null observation of $\tilde{\chi}$ in the CDMS
and XENON100 experiments has set  rather tight upper limits on the
spin-independent (SI) cross section of $\tilde{\chi}$-nucleon
scattering \cite{CDMSII, XENON100}. The CoGeNT experiment
\cite{CoGeNT} reported an excess which cannot be explained by any
known background sources but seems to be consistent with the signal
of a light $\tilde{\chi}$ with mass around 10 GeV and scattering
rate $\hbox{(1--2)} \times 10^{-40}$ cm$^2$.
Intriguingly, this range of mass and scattering rate are compatible
with dark matter explanation for both the DAMA/LIBRA data and the preliminary CRESST
data \cite{Hooper}. Though CoGeNT result is not consistent with the
CDMS or XENON results, it implies that the mass of dark matter can
range a very long scope at EW scale, that is from a few GeV to
several TeV. The indirect detection PAMELA also observed an excess
of the cosmic ray positron in the energy range 10-100 GeV ~\cite{pamela}
which may be explained by dark matter.

In this paper, We will give a short review on the difference
between the MSSM and the MSSM with a singlet sector under the constraints of
new dark matter detection results. As the Higgs hunting on colliders
has delicate relation with dark matter detections, the implication
on Higgs searching is also reviewed. The content is
based on our previous work \cite{Wang:2009rj,Cao:2010fi,Cao:2011re}.
the paper is organized as following, in sec. \ref{sec2},
we will give a short review on the structures of the MSSM, the NMSSM and the nMSSM.
In sec. \ref{sec3}  we will give a comparison on the models
under the constraints of CDMS, XENON, and CoGeNT. In sec. \ref{sec4},
a general singlet extension of the MSSM is discussed, and a
summary is given in sec. \ref{sec5}.

\section{the MSSM and its Singlet Extensions}
\label{sec2}
As the economical realization of supersymmetry, the MSSM has the
minimal content of particles, while the NMSSM and the nMSSM extend the
MSSM by only adding one singlet Higgs superfield $\hat{S}$. The
difference between these models is reflected in their
superpotential:
\begin{eqnarray}
W_{\rm MSSM}  & = & W_F+\mu \hat H_u\cdot\hat H_d,\\
W_{\rm NMSSM} & = & W_F + \lambda \hat{H}_u\cdot\hat{H}_d \hat{S}
                  +\frac{1}{3} \kappa  \hat{S}^3, \\
W_{\rm nMSSM} & = &  W_F + \lambda \hat{H}_u\cdot\hat{H}_d \hat{S}
                  + \xi_F M_n^2 \hat{S},
\label{sptial}
\end{eqnarray}
where
$W_F= Y_u  \hat{Q}\cdot\hat{H}_u  \hat{U}
    -Y_d \hat{Q}\cdot\hat{H}_d \hat{D}
    -Y_e \hat{L}\cdot\hat{H}_d \hat{E}$
with $\hat{Q}$, $\hat{U}$ and $\hat{D}$ being the squark
superfields, and $\hat{L}$ and $\hat{E}$ being the slepton
superfields. $\hat{H}_u$ and $\hat{H}_d$ are the Higgs doublet
superfields,  $\lambda$, $\kappa$ and $\xi_F$ are dimensionless
coefficients, and $\mu$ and $M_n$ are parameters with mass
dimension. Note that there is no explicit $\mu$-term in the NMSSM or
the nMSSM, and an effective $\mu$-parameter (denoted as $\mu_{\rm eff}$)
can be generated when the scalar component ($S$) of $\hat{S}$ develops
 a VEV. Also note that the nMSSM differs from the NMSSM
in the last term with the trilinear singlet term $\kappa \hat{S}^3$
of the NMSSM replaced by the tadpole term $\xi_F M_n^2 \hat{S}$. As
pointed out in Ref. \cite{xnMSSM1}, such a tadpole term can be generated
at a high loop level and naturally be of the SUSY breaking scale.
The advantage of such replacement is the nMSSM has no discrete
symmetry thus free of the domain wall problem which the NMSSM　
suffers from.

Corresponding to the superpotential, the Higgs soft terms in the
scalar potentials are also different between the three models (the soft
terms for gauginos and sfermions are the same thus not listed here)
 \begin{eqnarray}
V_{\rm soft}^{\rm MSSM}&=&\tilde m_{d}^{2}|H_d|^2 + \tilde m_{u}^{2}|H_u|^2
            + \left( B\mu H_u\cdot H_d +  \mbox{h.c.} \right) \\
V_{\rm soft}^{\rm NMSSM}&=&\tilde m_{d}^{2}|H_d|^2 + \tilde m_{u}^{2}|H_u|^2
            + \tilde m_s^{2}|S|^2
           + \left( A_\lambda \lambda S H_d\cdot H_u
           + \frac{\kappa}{3} A_{\kappa} S^3 + \mbox{h.c.} \right) , \\
V_{\rm soft}^{\rm nMSSM} & = & \tilde{m}_d^2 |H_d|^2 + \tilde{m}_u^2 |H_u|^2
   + \tilde{m}_s^2 |S|^2
   + \left( A_\lambda \lambda S H_d\cdot H_u + \xi_S M_n^3 S
   + \mbox{h.c.} \right).
\label{soft}
\end{eqnarray}
After the scalar fields $H_u$,$H_d$ and $S$ develop their VEVs
$v_u$, $v_d$ and $s$ respectively, they can be expanded as
\begin{eqnarray}
H_d  = \left ( \begin{array}{c}
             \frac{1}{\sqrt{2}} \left( v_d + \phi_d + i \varphi_d \right) \\
             H_d^- \end{array} \right) \, ,
H_u  = \left ( \begin{array}{c} H_u^+ \\
       \frac{1}{\sqrt{2}} \left( v_u + \phi_u + i \varphi_u \right)
        \end{array} \right)  \, ,
S  = \frac{1}{\sqrt{2}} \left( s + \sigma + i \xi \right)  \, .
\end{eqnarray}
The mass eigenstates can be obtained by unitary rotations
\begin{eqnarray}
\left( \begin{array}{c} h_1 \\ h_2\\ h_3\end{array} \right)
= U^H \left( \begin{array}{c} \phi_d \\ \phi_u \\ \sigma \end{array} \right),~
\left(\begin{array}{c} a_1 \\ a_2 \\ G_0\end{array} \right)
= U^A \left(\begin{array}{c}\varphi_d \\ \varphi_u\\ \xi \end{array} \right),~
 \left(\begin{array}{c}G^+\\ H^+ \end{array} \right)
=U^{H^+} \left(\begin{array}{c} H_d^+ \\ H_u^+\end{array}  \right),\label{rotation}
\end{eqnarray}
where $h_{1,2,3}$ and $a_{1,2}$ are respectively the CP-even and CP-odd
neutral Higgs bosons, $G^0$ and $G^+$ are Goldstone bosons, and $H^+$ is
the charged Higgs boson. Including the scalar part of the singlet sector
in the NMSSM and the nMSSM leads to a pair of charged Higgs bosons,
three CP-even and two CP-odd neutral Higgs bosons.
In the MSSM, we only have two CP-even and one CP-odd
neutral Higgs bosons in addition to a pair of charged Higgs bosons.

The MSSM predicts four neutralinos  $\chi^0_i$ ($i=1,2,3,4$), i.e.
the mixture of neutral gauginos (bino $\lambda'$ and neutral wino
$\lambda^3$) and neutral higgsinos ($\psi_{H_u}^0, \psi_{H_d}^0$),
while the NMSSM and the nMSSM predict one more neutralino corresponding
to the singlino $\psi_S$ from the fermion part of singlet sector.
In the basis $(-i\lambda',- i \lambda^32, \psi_{H_u}^0, \psi_{H_d}^0, \psi_S )$
(for the MSSM $\psi_S$ is absent) the  neutralino mass matrix is given by
\begin{eqnarray}\label{mass-matrix1}
\left( \begin{array}{ccccc}
M_1          & 0             & m_Zs_W s_b    & - m_Z s_W c_b  & 0 \\
0            & M_2           & -m_Z c_W s_b  & m_Z c_W c_b    & 0 \\
m_Zs_W s_b   & -m_Z s_W s_b  & 0             & -\mu           & -\lambda v c_b \\
-m_Z s_W c_b & -m_Z c_W c_b  &  -\mu         & 0              & - \lambda v s_b \\
0            & 0             &-\lambda v c_b &- \lambda v s_b &
\scriptstyle\left\{\begin{array}{c}
    \scriptstyle 2\frac{\kappa}{\lambda}\mu~~{\rm{~~for ~the~ NMSSM}}\\
\scriptstyle  0 {\rm~~~~~~for ~the~ nMSSM}
\end{array}\right.
\end{array} \right),
\end{eqnarray}
where $M_1$ and $M_2$ are respectively $U(1)$ and $SU(2)$ soft gaugino
mass parameters,
$s_W=\sin \theta_W$, $c_W=\cos\theta_W$, $s_b=\sin\beta$
and $c_b=\cos\beta$
with $\tan \beta \equiv v_u/v_d$.
The lightest neutralino  $\tilde{\chi}^0_1$ is assumed to
be the lightest supersymmetric particle (LSP),
serving as the SUSY dark matter
particle. It is composed by
\begin{eqnarray}
\tilde{\chi}^0_1=N_{11}(-i\lambda')+N_{12}(-i \lambda^3)
+N_{13}\psi_{H_u}^0 +N_{14} \psi_{H_d}^0 + N_{15}\psi_S,\label{singlino}
\end{eqnarray}
where $N$ is the unitary matrix ($N_{15}$ is zero for the MSSM)
to diagonalize the mass matrix in Eq. (\ref{mass-matrix1}).
For the mass matrices above we should note that the following two points
\begin{enumerate}
\item For a moderate value of $\kappa$, the neutralino sector of the NMSSM
can go back to the MSSM when  $\lambda$ approaches to zero.
This is because in such case
the singlino component will become super heavy and decouple
from EW scale. The singlet scalar will not mix with the two Higgs doublet,
then the NMSSM will be almost the same as the MSSM at EW scale.
\item Since the $\psi_S\psi_S$ element of Eq. (\ref{mass-matrix1}) is zero
in the nMSSM, the singlino will not decouple when $\lambda$ approaches to zero.
 In fact,  in the nMSSM the mass of the LSP can be written as
\begin{eqnarray}
m_{\chi_1^0} \simeq \frac{2 \mu_{\rm eff} \lambda^2 ( v_u^2 + v_d^2
)}{2 \mu_{\rm eff}^2 + \lambda^2 (v_u^2 + v_d^2 )} \frac{\tan \beta}{\tan^2
\beta + 1 }.\label{m_chi0}
\end{eqnarray}
This formula shows that to get a heavy $\tilde{\chi}^0_1$, we need a
large $\lambda$, a small $\tan \beta$ as well as a moderate $\mu_{\rm eff}$.
\end{enumerate}

The chargino sector of these three models is the same except
that in the NMSSM/nMSSM the parameter $\mu$ is replaced by $\mu_{\rm eff}$.
The charginos $\tilde{\chi}^\pm_{1,2}$ ($m_{\chi^\pm_1}\leq m_{\chi^\pm_2}$)
are the mixture of charged Higgsinos $\psi_{H_{u,d}}^\pm$
and winos $\lambda^\pm=(\lambda^1\pm\lambda^2)/\sqrt{2}$, whose mass
matrix in the basis of  $(-i\lambda^\pm,\psi_{H_{u,d}}^\pm)$ is given by
\begin{eqnarray}
\left( \begin{array}{cc}
 M_2 &             \sqrt{2} m_W s_b \\
\sqrt{2} m_W c_b & \mu_{\rm eff}
\end{array} \right).
\end{eqnarray}
So the chargino $\tilde{\chi}^\pm_1$ can be wino-dominant (when $M_2$ is much smaller than
$\mu$) or higgsino-dominant (when $\mu$ is much smaller than $M_2$).
Since the composing property (wino-like, bino-like, higgsino-like or singlino-like)
of the LSP and the chargino $\tilde{\chi}^\pm_1$ is very important in SUSY phenomenologies,
we will show such a property in our following study.

\section{Comparison with the MSSM and  the MSSM with a  Singlet sector}\label{sec3}
\subsection{In light of CDMS II and XENON}
First let's see the MSSM, the NMSSM and the nMSSM under the constraints of results of
CDMS II and XENON100. As both current and future limits of $\tilde\chi$-nucleon
of CDMS and XENON are similar to each other,  we will show only one of them.
Nevertheless, as a good substitute of the SM, SUSY model must satisfy all the
results of current collider and detector measurements.
In our study we consider the following
experimental constraints: \cite{Nakamura:2010zzi}
(1) we require $\tilde{\chi}^0_1$ to account for
dark matter relic density $0.105 < \Omega h^2 < 0.119$; (2) we
require the SUSY contribution to explain the deviation of the muon
$a_\mu$, i.e., $a_\mu^{\rm exp} - a_\mu^{\rm SM} = ( 25.5 \pm 8.0 ) \times
10^{-10}$, at $2 \sigma$ level; (3) the LEP-I bound on the invisible
$Z$-decay, $\Gamma(Z \to \tilde{\chi}^0_1 \tilde{\chi}^0_1) < 1.76$
MeV, and the LEP-II upper bound on $\sigma(e^+e^- \to
\tilde{\chi}^0_1 \tilde{\chi}^0_i)$, which is $5 \times 10^{-2}~{\rm
pb}$ for $i>1$, as well as the lower mass bounds on sparticles from
direct searches at LEP and the Tevatron; (4) the constraints from
the direct search for Higgs bosons at LEP-II, including the decay
modes $h \to h_1 h_1, a_1 a_1 \to 4 f$, which limit all possible
channels for the production of the Higgs bosons; (5) the constraints
from $B$ physics observable such as $B \to X_s \gamma$, $B_s \to
\mu^+\mu^-$, $B^+ \to \tau^+ \nu$, $\Upsilon \to \gamma a_1 $, the
$a_1$--$\eta_b$ mixing and the mass difference $\Delta M_d$ and $\Delta
M_s$; (6) the constraints from the precision EW observable
such as $\rho_{\rm lept}$, $\sin^2 \theta_{\rm eff}^{\rm lept}$, $m_W$ and
$R_b$; (7) the constraints from the decay $\Upsilon \to \gamma h_1$,
and the Tevatron search for a light Higgs boson via $4 \mu$ and
$2 \mu 2 \tau$ signals \cite{Dark-Higgs}. The constraints (1--5) have
been encoded in the package NMSSMTools \cite{NMSSMTools}. We use
this package in our calculation and extend it by adding the
constraints (6, 7). As pointed out in Ref. \cite{Dark-Higgs}, the
constraints (7) are important for a light Higgs boson.
In addition to the above experimental limits, we also consider the
constraint from the stability of the Higgs potential, which requires
that the physical vacuum of the Higgs potential with non-vanishing
VEVs of Higgs scalars should be lower than any local minima.

For the calculation of cross section of $\tilde\chi$-nucleon
scattering, we use the formulas in Ref. \cite{Drees,susy-dm-review} for
the MSSM and extend them to the NMSSM/nMSSM. It is sufficient to
consider only the SI interactions between $\tilde{\chi}^0_1$ and
nucleon (denoted by $f_p$ for proton and $f_n$ for neutron
\cite{susy-dm-review}) in the calculation.
The leading order of these interactions are induced by exchanging the SM-like Higgs
boson at tree level. For moderately light Higgs bosons,
 $f_p$ is approximated by \cite{susy-dm-review} (similarly for$f_n$)
\begin{equation}
 \begin{split}
    f_{p}  \simeq
\sum_{q=u, d, s} \frac{f_q^{H}}{m_q} m_p f_{T_q}^{(p)}
    + \frac{2}{27}f_{T_G} \sum_{q=c, b, t} \frac{f_q^{H}}{m_q} m_p~,
 \end{split}     \label{2b}
\end{equation}
where $f_{Tq}^{(p)}$ denotes the fraction of $m_p$ (proton mass)
from the light quark $q$ while
$f_{T_G}=1-\sum_{u,d,s}f_{T_q}^{(p)}$ is the heavy quark
contribution through gluon exchange. $f_q^{H}$ is the
coefficient of the effective scalar operator. The
$\tilde \chi^0$-nucleus scattering rate is then given by
\cite{susy-dm-review}
\begin{equation}
    \sigma^{SI} = \frac{4}{\pi}
    \left( \frac{m_{\tilde \chi^0} m_T}{m_{\tilde \chi^0} + m_T} \right)^2
    \times \bigl( n_p f_p + n_n f_n \bigr)^2,
\end{equation}
where $m_T$ is the mass of target nucleus and $n_p (n_n)$ is the number of
proton (neutron) in the target nucleus.
In our numerical calculations we take
$f_{T_u}^{(n)} =0.023$, $f_{T_d}^{(n)} = 0.034$, $f_{T_u}^{(p)} = 0.019$,
$f_{T_d}^{(p)} = 0.041$ and $f_{T_s}^{(p)} = f_{T_s}^{(n)} = 0.38$.
Note that the scattering rate is very sensitive to the value of
$f_{T_s}$ \cite{Ellis:2008hf}. Recent lattice simulation \cite{lattice}
gave a much smaller value of $f_{T_s}$ (0.020), it reduces the
scattering rate significantly which can be seen in Ref. \cite{Cao1}.

Considering all the constraints listed above,
we scan over the parameters in the following ranges
\begin{eqnarray}
&& 100 {\rm ~GeV} \leq
\left(M_{\tilde{q}},M_{\tilde{\ell}}, ~m_A, ~\mu \right) \leq  1 {\rm ~TeV},
\nonumber\\
&& 50 {\rm ~GeV} \leq M_1 \leq  1 {\rm ~TeV}, ~~1 \leq  \tan \beta \leq 40, \nonumber\\
&& \left( |\lambda|,|\kappa| \right) \leq 0.7, ~~|A_\kappa| \leq  1 {\rm ~TeV},\label{range}
\end{eqnarray}
where $M_{\tilde{q}}$ and $M_{\tilde{\ell}}$ are the universal soft mass
parameters of the first two generations of squarks and the three generations
of sleptons respectively. To reduce the number of the relevant soft parameters,
we worked in the so-called $m_h^{max}$ scenario with following choice of the soft
masses for the third generation squarks:
$M_{\tilde Q_3}=M_{\tilde U_3}=M_{\tilde D_3}=800$ GeV, and $X_t = A_t - \mu \cot \beta =
-1600$ GeV.  The advantage of such a choice is that other SUSY
parameters more easily survive the constraints (so that the bounds
we obtain are conservative). Moreover, we assume
the grand unification relation for the gaugino masses:
\begin{equation}
 M_1:M_2:M_3\simeq 1:1.83:5.26~. \label{gut_relation}
\end{equation}
This relation is often assumed in studies of SUSY at the
 TeV scale for it can be easily generated in the mSUGRA model \cite{Nilles:1983ge}.
Note that relaxing this relation will give  a large effect on
the light neutralino scenario \cite{Feldman:2010ke}.

 \begin{figure}[htbp]
 \epsfig{file=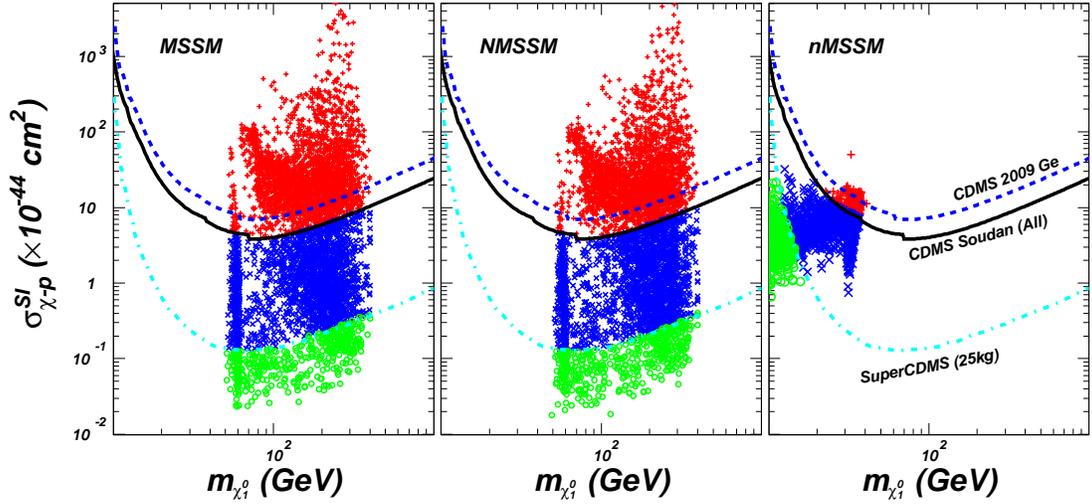,height=7cm}
 \vspace{-0.3cm}
\caption{The scatter plots (taken for Ref. \cite{Cao:2010fi})
for the spin-independent elastic cross
section of $\tilde\chi$-nucleon scattering. The `$+$' points (red) are
excluded by CDMS limits (solid line),  the `$\times$' (blue) would
be further excluded by SuperCDMS 25kg \cite{supercdms}
in case of non-observation (dash-dotted line), and the `$\circ$'
(green) are beyond the SuperCDMS sensitivity.}
   \label{fig1}
\end{figure}
The surviving points for the three model are displayed in Fig.~\ref{fig1}
for the spin-independent elastic cross section of $\tilde\chi$-nucleon scattering.
We see that for each model the CDMS II limits can exclude a large
part of the parameter space allowed by current collider constraints
and the future SuperCDMS (25 kg) limits can cover most of the allowed
parameter space. For the MSSM and the NMSSM  dark matter mass is
roughly in range of 50-400 GeV, while for the nMSSM  dark
matter mass is constrained below 40 GeV by current experiments and
further constrained below 20GeV by SuperCDMS in case of
non-observation.

\begin{figure}[htbp]
 \epsfig{file=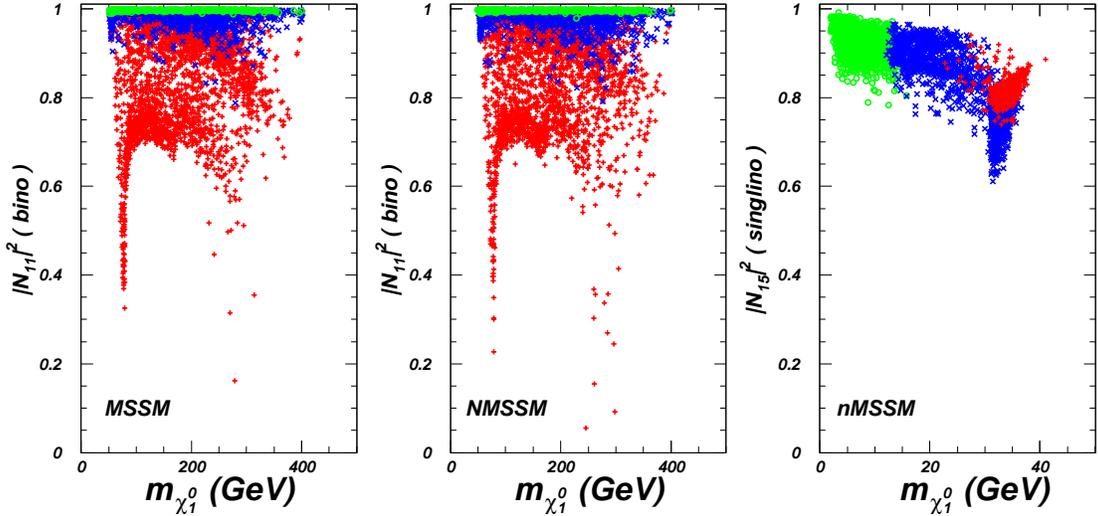,height=7cm}
 \vspace{-0.3cm}
\caption{Same as Fig.~\ref{fig1}, but projected on the plane of
         $|N_{11}|^2$ and $|N_{15}|^2$ versus dark matter mass.
         (taken for Ref. \cite{Cao:2010fi})}
\label{fig2}
\end{figure}
\begin{figure}[htbp]
\epsfig{file=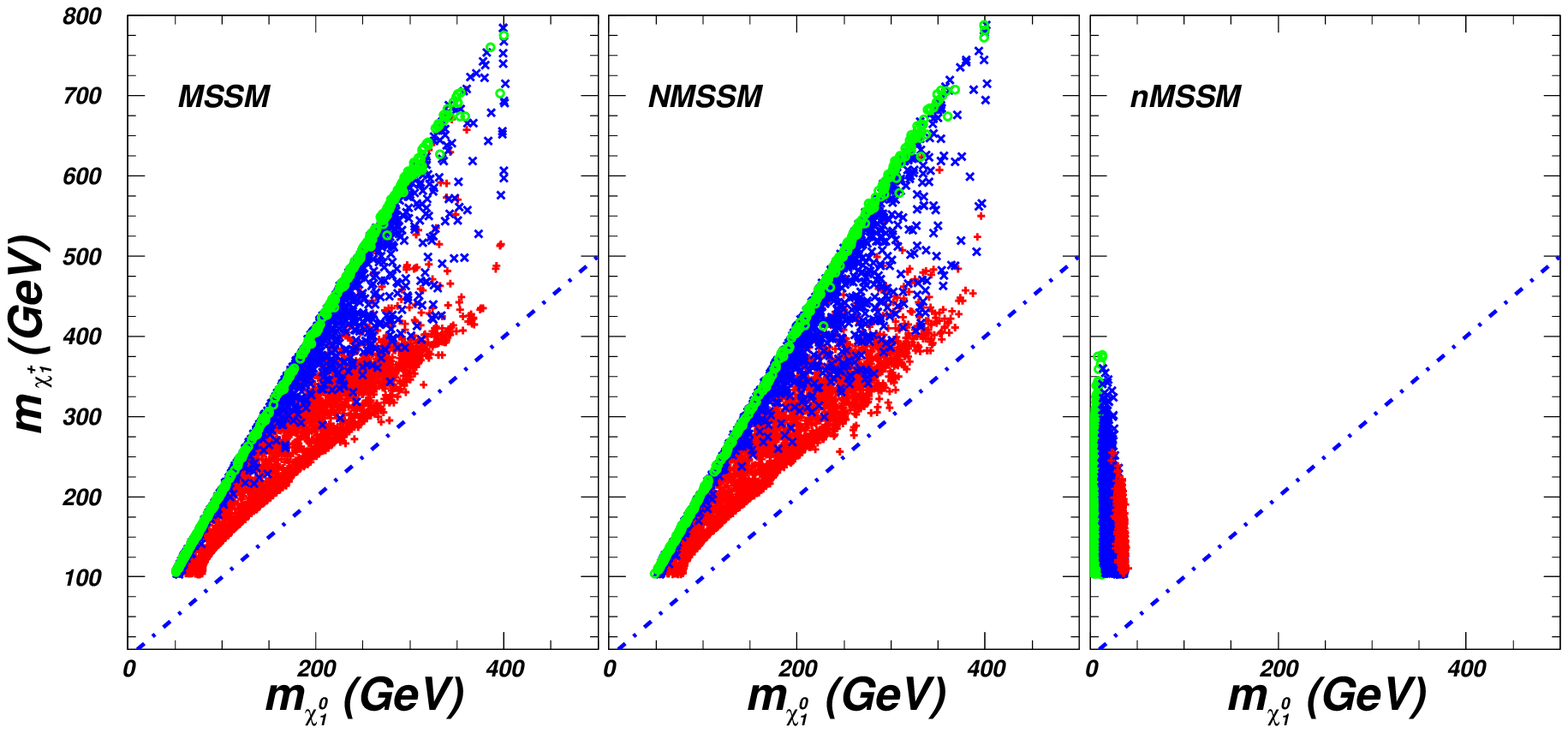,height=7cm} \vspace{-0.3cm} \caption{Same as
Fig.~\ref{fig1}, but showing
         the chargino mass $m_{\chi^+_1}$ versus the LSP mass.
         The dashed lines indicate $m_{\chi^+_1}=m_{\chi^0_1}$.
(taken for Ref. \cite{Cao:2010fi})}
\label{fig3}
\end{figure}
From Fig.~\ref{fig1}, we can see that the $\tilde\chi$-nucleon scattering plot
of the MSSM and the NMSSM are very similar to each other, but very different
from nMSSM. This implies that under the experiment constraints,
the singlet sector will decouple
from the MSSM-like sector in the NMSSM, then the NMSSM will perform almost
 the same as the MSSM, However,
the singlet components change EW scale phenomenology greatly in the nMSSM.
This can also be seen in Fig.~\ref{fig2} and Fig.~\ref{fig3}.
We can see that for both the MSSM and the NMSSM $\tilde{\chi}_1^0$ is bino-dominant,
while for the nMSSM $\tilde{\chi}_1^0$ is singlino-dominant, and the region
allowed by CDMS limits (and SuperCDMS limits in case of
non-observation) favors a more bino-like $\tilde{\chi}_1^0$ for the  MSSM/NMSSM
and a more singlino-like $\tilde{\chi}_1^0$ for the nMSSM.
For the MSSM/NMSSM the LSP lower bound around 50 GeV
is from the chargino lower bound of 103.5 GeV plus the assumed GUT
relation $M_1 \simeq 0.5 M_2$; while the upper bound around 400 GeV
is from the bino nature of the LSP ($M_1$ cannot be too large, must
be much smaller than other relevant parameters) plus the
experimental constraints like the muon g-2 and B-physics. If we do
not assume the GUT relation $M_1 \simeq 0.5 M_2$, then $M_1$ can be
as small as 40 GeV and the LSP lower bound in the MSSM/NMSSM will not be
sharply at 50 GeV. (We talk about it in the following section.)
For both the MSSM and the NMSSM, the CDMS limits tend to favor a heavier
chargino and ultimately the SuperCDMS limits tend to favor a
wino-dominant chargino with mass about $2 m_{\chi^0_1}$.
Note that, there still can be a singlino dominant LSP in
some parameter space of the NMSSM \cite{Belanger:2005kh},
but in the scan range Eq. (\ref{range}) listed above,  getting
such singlino dominant LSP needs some fine-tuning,
thus we do not focus on it.
\begin{figure}[htbp]
\epsfig{file=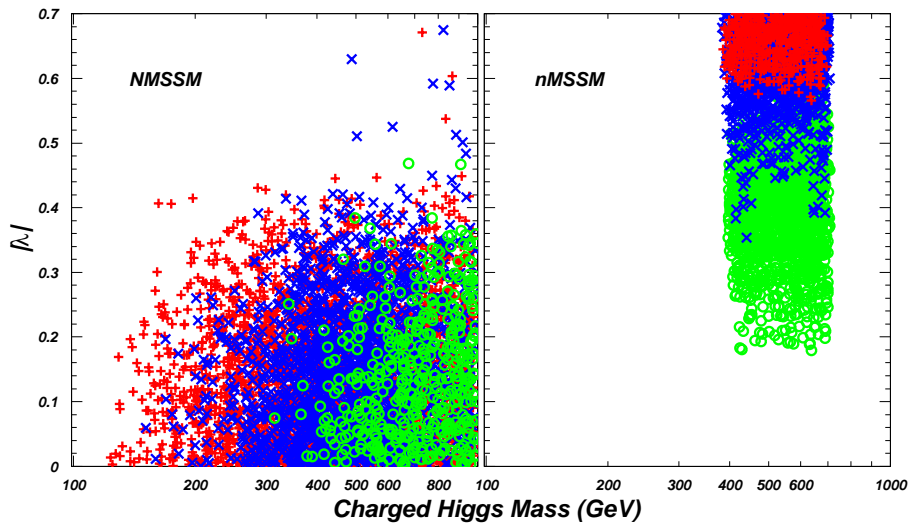,height=7cm} \vspace{-0.3cm} \caption{Same as
Fig.~\ref{fig1}, but projected on the plane of
         $|\lambda|$ versus the charged Higgs mass in the NMSSM and the nMSSM.
(taken for Ref. \cite{Cao:2010fi})}
\label{fig4}
    \end{figure}

In Fig.~\ref{fig4} we show the value of $|\lambda|$ versus the
charged Higgs mass in the NMSSM and the nMSSM. This figure indicates that
$\lambda$ larger than 0.4 is disfavored by the NMSSM. The
underlying reason is that $h_1 \tilde{\chi}^0_1\tilde{\chi}^0_1$ depends on
$\lambda$ explicitly and large $\lambda$ can enhance $\tilde\chi$-nucleon
scattering rate.  By contrast,
although CDMS has excluded some points with large $\lambda$ in
the nMSSM, there are still many surviving points with $\lambda$
as large as 0.7. We have talked the reason above:
to get a heavy $\chi^0_1$, one need a large $\lambda$,
a small $\tan \beta$ as well as a moderate $\mu_{\rm eff}$.

\begin{figure}[htbp]
\epsfig{file=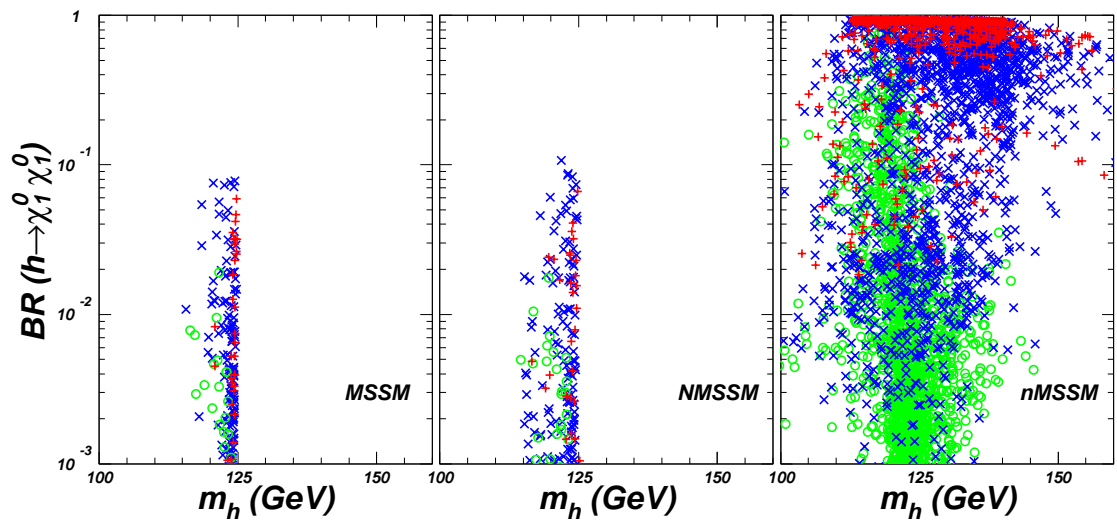,height=7cm} \vspace{-0.3cm} \caption{Same as
Fig.~\ref{fig1}, but projected for
 the decay branching ratio of $h_{\rm SM} \to \chi^0_1 \chi^0_1$
 versus the mass of the Higgs boson $h_{\rm SM}$.
(taken for Ref. \cite{Cao:2010fi})}
\label{fig5}
    \end{figure}
From the survived parameter space for all the model above,
we should know that the Higgs decay will be similar for the MSSM
and the NMSSM, but quite different from the nMSSM. This can be seen in
Fig.~\ref{fig5} which shows decay branching ratio of $h_1 \to
\tilde{\chi}^0_1 \tilde{\chi}^0_1$ versus the mass of the
SM-like Higgs boson $h_{\rm SM}$ ( which is $h_1$ here, and it
is Higgs doublet $\hat{H}_u$ and $\hat{H}_d$ dominant ).
Such a decay is strongly correlated to the $\tilde\chi$-nucleon scattering
because the coupling $h_1 \tilde{\chi}^0_1\tilde{\chi}^0_1$ is involved in both
processes. We see that in the MSSM and the NMSSM this decay mode can
open only in a very narrow parameter space since $\tilde{\chi}_1^0$ cannot
be so light, and in the allowed region this decay has a very small
branching ratio (below $10\%$). However, in the nMSSM this decay
can open in a large part of the parameter space since the LSP can be
very light, and its branching ratio can be quite large (over $80\%$
or  $90\%$).

\subsection{light dark matter in the NMSSM}
As talked in the introduction, the data of CoGeNT experiment
favors  a light dark matter  around 10 GeV.
However,  we scan the parameter space in the MSSM  and find
that it is very difficult to find a neutralino  $\tilde{\chi}_1^0$
lighter than about 28 GeV, unless when it is associated with a light stau as
the next to the lightest supersymmetric particle (NLSP),
but such scenario always needs a fine-tuning
in the parameter space \cite{Dreiner:2009ic}.
The main reason for the absence of a lighter
$\tilde{\chi}_1^0$ is that the dominant annihilation channel for
$\tilde{\chi}_1^0$ in the early universe is
$\tilde{\chi}_1^0\tilde{\chi}_1^0 \to b\bar{b}$ through
$s$-channel exchange of the pseudoscalar Higgs boson ($A$) and the
measured dark matter relic density requires $m_A \sim (90\endash100)$ GeV
and $\tan \beta \sim 50$, this is in conflict with the constraints
from the LEP experiment and $B$ physics
\cite{MSSM-light,NMSSM-light,Belanger}. The LHC data
gives an even more stronger constraint on the
light pseudoscalar scenario \cite{Dermisek:2009fd}
such that light dark matter seems impossible in the MSSM.
Though in the nMSSM the neutralino $\tilde{\chi}_1^0$ can be as
light as 10 GeV (shown in Fig.~\ref{fig1}),
the scattering rate is much lower under the CoGeNT-favored region.
In the NMSSM, however, with the participation of singlet sector one can get
very light \cite{NMSSM} Higgs. This feature is
particularly useful for light $\tilde{\chi}_1^0$ scenario
since it opens up new important annihilation channels for
$\tilde{\chi}_1^0$, i.e., either into a pair of $h_1$ (or $a_1$) or
into a pair of fermions via $s$-channel exchange of $h_1$ (or
$a_1$) \cite{Belanger,light-anni,Cao-nMSSM}. For the former case,
$\tilde{\chi}_1^0$ must be heavier than $h_1$ ($a_1$); while for the
latter case, due to the very weak couplings of
 $h_1$ ($a_1$) with $\tilde{\chi}_1^0$ and with the SM fermions,
a resonance enhancement (i.e.  $m_{h_1}$ or  $m_{a_1}$ must be close
to $2m_{\tilde{\chi}_1^0}$) is needed to accelerate the
annihilation. So a light $\tilde{\chi}_1^0$ may be necessarily
accompanied by a light $h_1$ or $a_1$ to provide the required dark
matter relic density. From the discussion in the upper section,
light $\tilde{\chi}_1^0$
can be obtained by releasing the GUT relation Eq. (\ref{gut_relation}),
thus LSP in the NMSSM may explain the detection of CoGeNT.
Note that, as the LSP in the  nMSSM is singlino dominant, relaxing the
GUT relation will not the change the phenomenology of dark matter
and Higgs too much.

\begin{figure}[htbp]
\includegraphics[width=13cm]{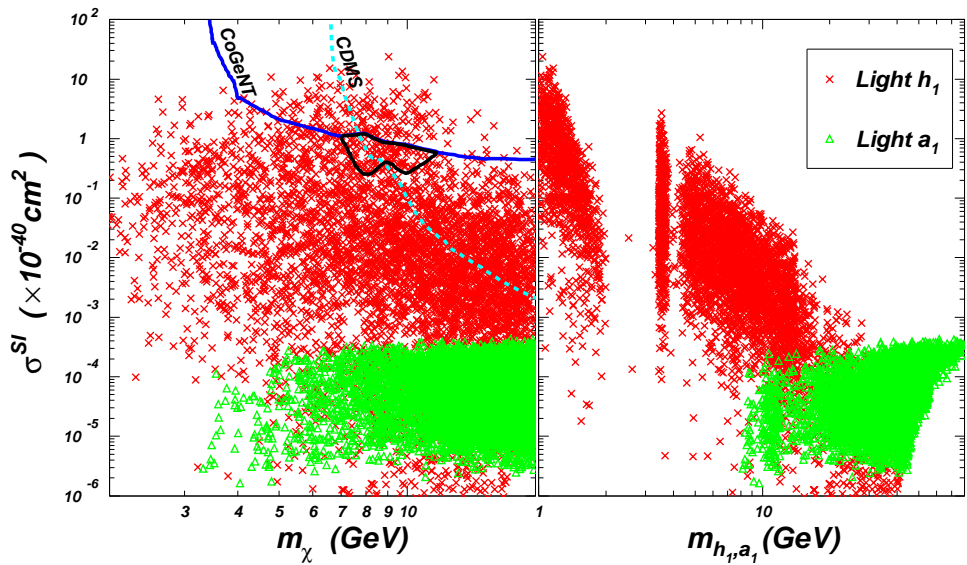}
\vspace{-0.5cm} \caption{The scatter plots (taken for Ref. \cite{Cao:2011re})
of the parameter samples
which survive all constraints, with `$\times$' (red)  and
`$\blacktriangle$' (green) corresponding to a light $h_1$ and a
light $a_1$, respectively. The left frame is projected on the
$\sigma^{\rm SI}$-$m_\chi$ plane, while the right frame is projected on
the $\sigma^{\rm SI}$-$m_{h_1}$ plane (denoted by `$\times$') and the
$\sigma^{\rm SI}$-$m_{a_1}$ plane (denoted by `$\blacktriangle$'). The
curves are the limits from CoGeNT \cite{CoGeNT}, CDMS \cite{CDMSII},
while the contour is the CoGeNT-favored region \cite{CoGeNT}.} \label{fig6}
\end{figure}
Now we discuss how to get  a light $h_1$ or $a_1$ in the NMSSM. A
light $a_1$ can be easily obtained when the theory is close to the
U(1)$_R$ or U(1)$_{\rm PQ}$ symmetry limit, which can be realized
by setting the product $\kappa A_\kappa$ to be negatively small
\cite{NMSSM}. In contrast, a light $h_1$ can not be obtained
easily. However, as shown below, it can still be achieved by somewhat
subtle cancelation via tuning the value of $A_\kappa$. We note
that for any theory with multiple Higgs fields, the existence of a
massless Higgs boson implies the vanishing of the determinant of
its squared mass matrix and vice versa. For the NMSSM, at tree
level the parameter $A_\kappa$ only enters the mass term of the
singlet Higgs bosons, so the determinant ($\Det{\cal{M}}^2$) of
the mass matrix of the CP-even Higgs bosons depends on $A_\kappa$
linearly \cite{NMSSM}. When other relevant parameters are fixed,
one can then obtain a light $h_1$ by varying $A_\kappa$ around the
value $\tilde{A}_\kappa$ which is the solution to the equation
$\Det{\cal{M}}^2=0$. In practice, one must include the important
radiative corrections to the Higgs mass matrix, which will
complicate the dependence of ${\cal{M}}^2$ on $A_\kappa$. However,
we checked that the linear dependence is approximately maintained
by choosing the other relevant parameters at the SUSY scale, and
one can solve the equation iteratively to get the solution
$\tilde{A}_\kappa$.

In Fig.~\ref{fig6} we display the surviving parameter samples,
showing the  $\tilde\chi$-nucleon scattering cross section
versus the neutralino dark matter mass (left frame) and versus the
mass of $h_1$ or $a_1$ (right frame). It shows that the scattering
rate of the light dark matter
can reach the sensitivity of CDMS and, consequently, a sizable parameter
space is excluded by the CDMS data \cite{supercdms}.
The future CDMS experiment can further
explore (but cannot completely cover) the remained parameter space.
Note that in the light-$h_1$ case the scattering rate can be
large enough to reach the sensitivity of CoGeNT and to cover the CoGeNT-favored
region. The underlying reason is that the $\tilde\chi$-nucleon scattering
can proceed through the $t$-channel exchange of the CP-even Higgs bosons,
which can be enhanced by a factor $1/m_{h_1}^4$ for a light $h_1$
\cite{light-anni}; while a light $a_1$ can not give such an
enhancement because the CP-odd Higgs bosons do not contribute to the
scattering in this way.  We noticed that the studies in
\cite{NMSSM-light,Das-light} claimed that the NMSSM is unable to
 explain the CoGeNT data because they did not consider the
light-$h_1$ case.

\begin{figure}[htb]
\includegraphics[width=14.0cm]{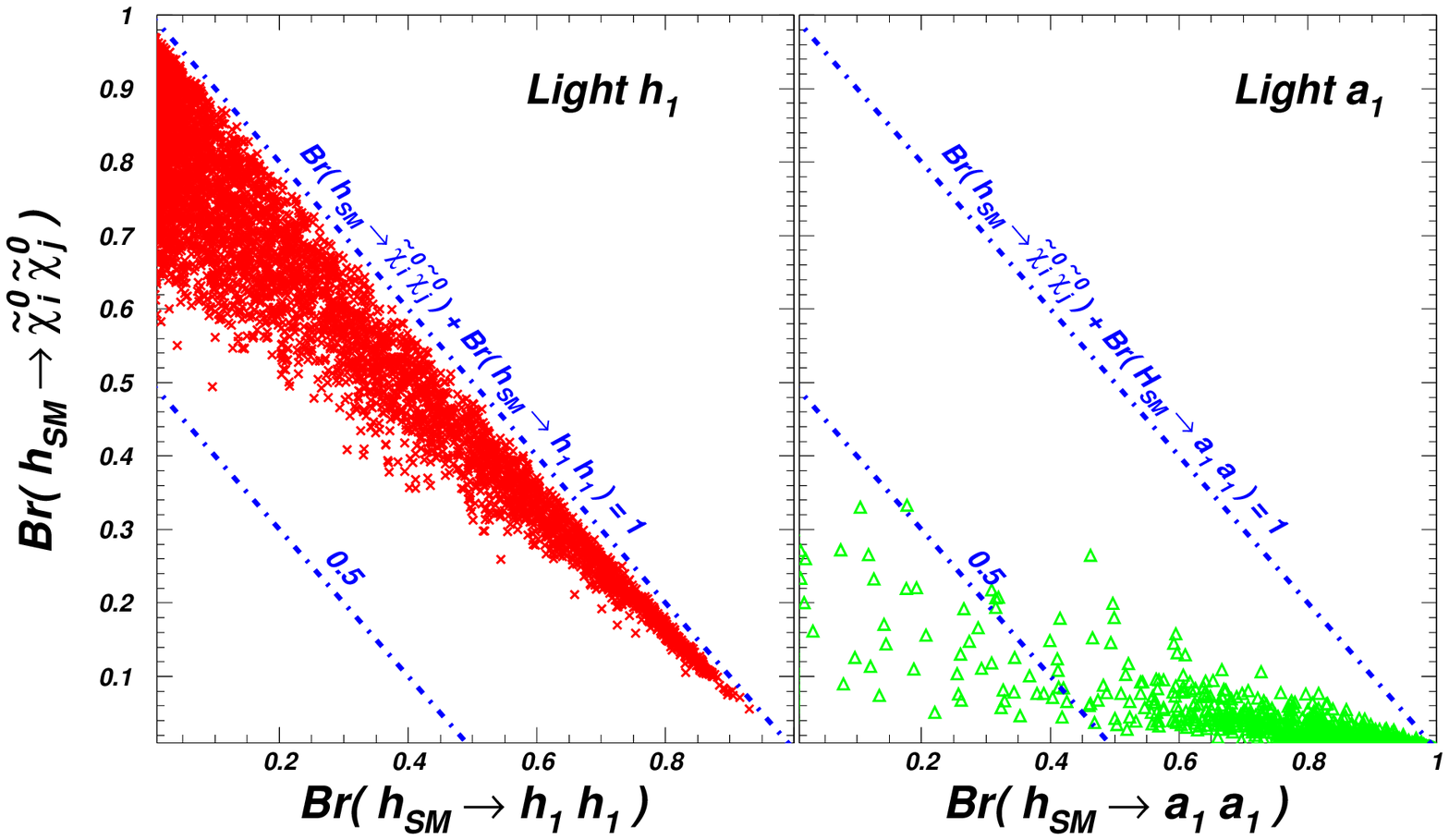} \hspace*{0.2cm}
\includegraphics[width=13.7cm]{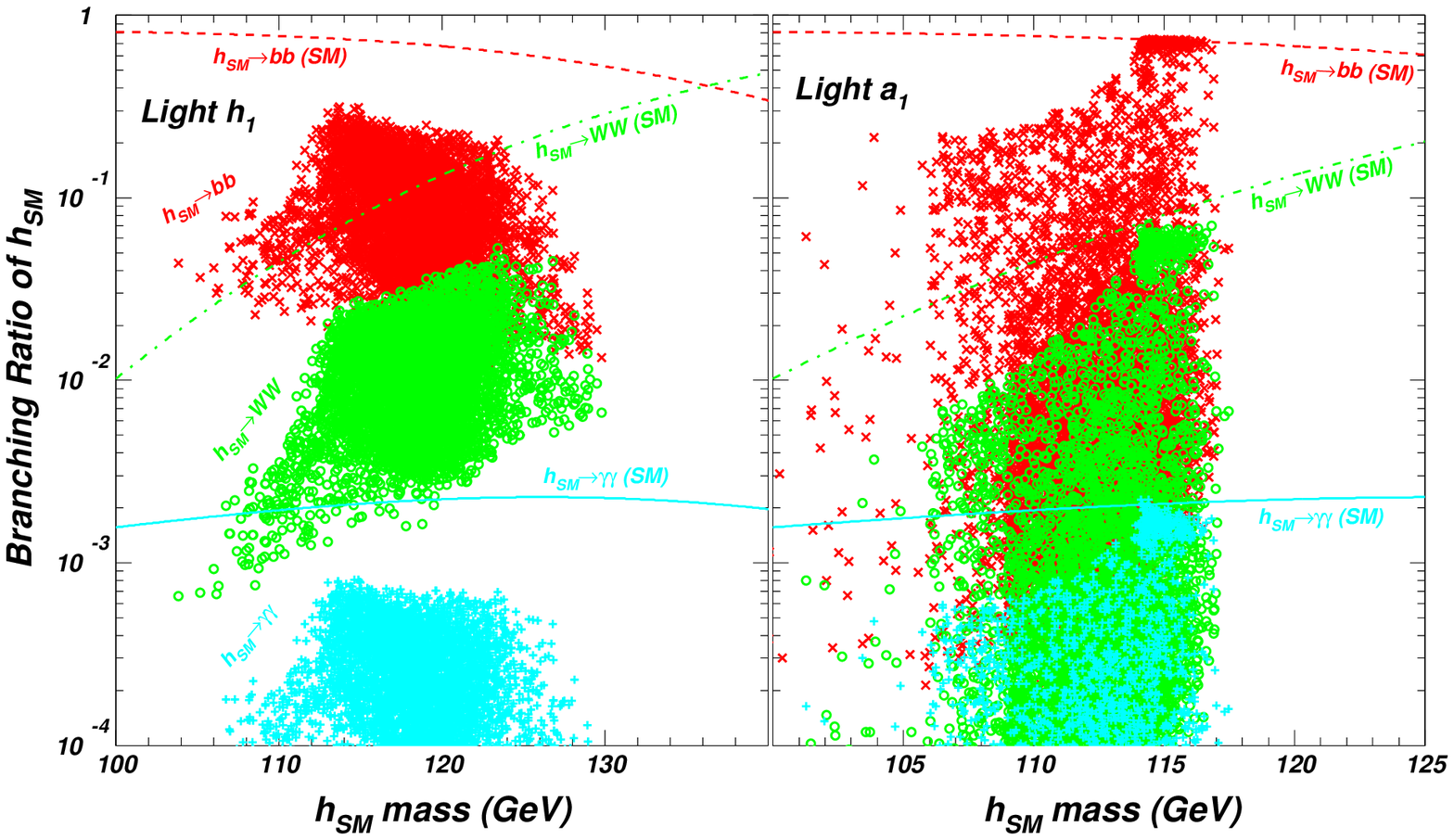}
\vspace{-0.8cm} \caption{Same as Fig. \ref{fig6}, but showing the
decay branching ratios of the SM-like Higgs boson $h_{\rm SM}$. Here
$Br(h_{\rm SM} \to \tilde{\chi}_i^0 \tilde{\chi}_j^0)$ denotes  the
total rates for all possible $h_{\rm SM} \to \tilde{\chi}_i^0
\tilde{\chi}_j^0$ decays. (taken for Ref. \cite{Cao:2011re})} \label{fig7}
\end{figure}
In the light $\tilde{\chi}_1^0$ scenario, $h_{\rm SM}$ may decay
exotically into $\tilde{\chi}_i^0 \tilde{\chi}_j^0$, $h_1 h_1$ or
$a_1 a_1$, and consequently the conventional decays are reduced.
This feature is illustrated in Fig.~\ref{fig7}, which shows that the
sum of the exotic decay branching ratios may exceed $50\%$ and the
traditional decays $h_{\rm SM} \to b \bar{b}, \tau \bar{\tau}, W
W^\ast, \gamma \gamma$ can be severely suppressed. Numerically,
we find that the branching ratio of $h_{\rm SM} \to b \bar{b}$
is suppressed to be below $30\%$ for all the surviving
samples in the light-$h_1$ ($h_2$ is $h_{\rm SM}$) case and for about $96\%$ of the
surviving samples in the light-$a_1$ ($h_1$ is $h_{\rm SM}$) case (for the remaining
$4\%$ of the surviving samples in the light-$a_1$ case,
the decay $h_{\rm SM} \to a_1 a_1$ is usually kinematically
forbidden so that the ratio of
$h_{\rm SM} \to b \bar{b}$ may exceed $60\%$).
Another interesting feature  shown in Fig.~\ref{fig7} is that, due
to the open-up of the exotic decays,  $h_{\rm SM}$ may be
significantly lighter than the LEP bound. This situation is
favored by the fit of the precision electro-weak data and is of
great theoretical interest \cite{Gunion}.

\begin{figure}[htb]
\includegraphics[width=9cm]{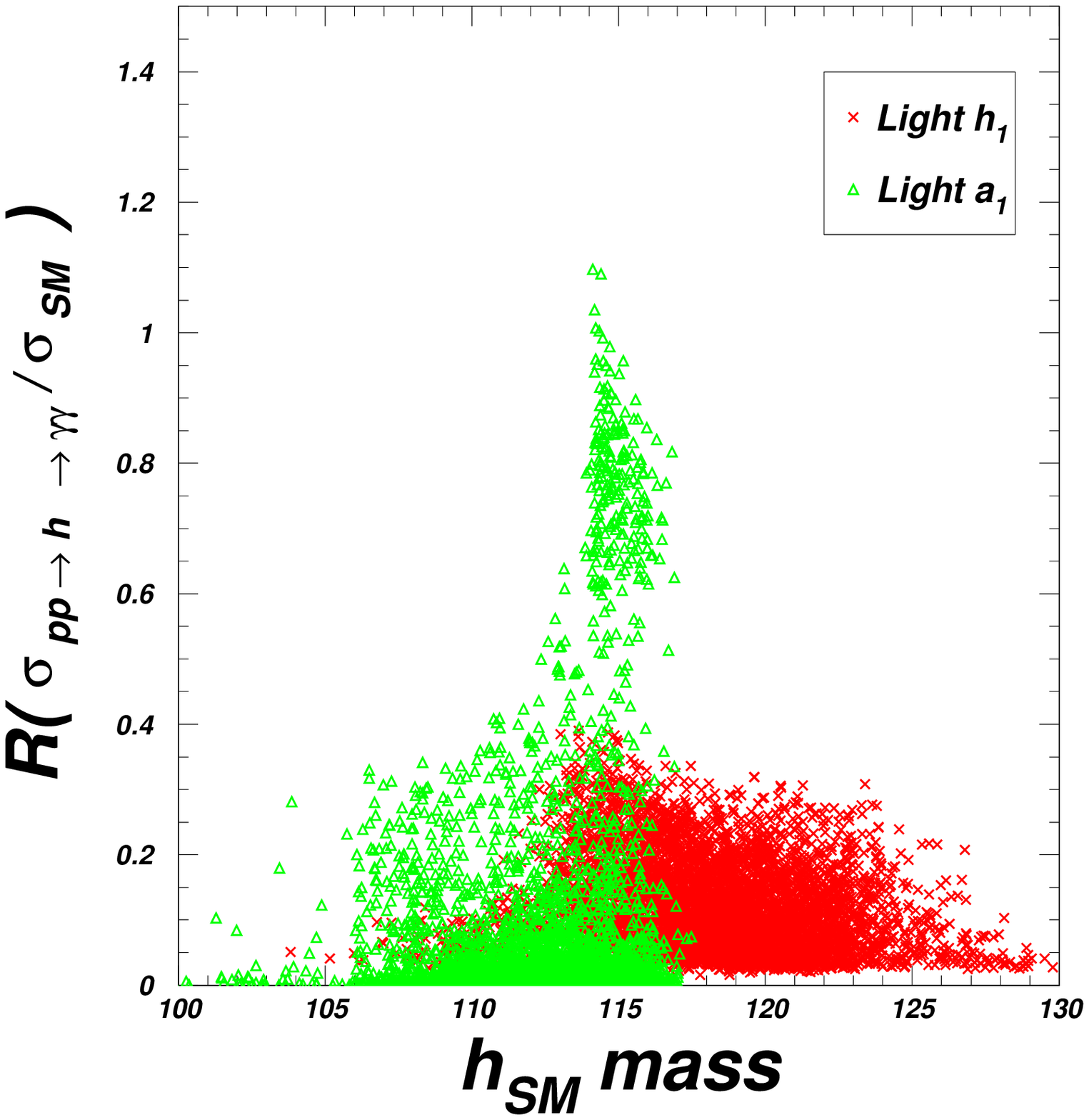}
\vspace{-0.8cm}
\caption{Same as Fig.~\ref{fig6}, but showing the
diphoton production rate of the SM-like Higgs boson at the LHC.}
\label{fig8}
\end{figure}
Since the conventional decay modes of $h_{\rm SM}$ may be greatly
suppressed, especially in the light-$h_1$ case which can give a
rather large $\tilde\chi$-nucleon scattering rate, the LHC search for
$h_{\rm SM}$ via the traditional channels may become difficult.
Now the LHC observed a new particle in the mass region around 125-126 GeV
which is the most probable the long sought Higgs boson \cite{cern}.
In this mass range,  the most important discovering
channel of $h_{\rm SM}$ at the LHC is the di-photon signal.
In Fig.~\ref{fig8} we give
the ratio of the di-photon  production rate to the SM at the LHC
with $\sqrt{s}=7$ TeV. In calculating the rate, we used
the narrow width approximation and only considered the
leading contributions to $p p \to h_{\rm SM}$ from
top quark, bottom quark and the squark loops.

Fig.~\ref{fig8} indicates that, compared with the SM prediction, the ratio
in the NMSSM  in the light $\tilde{\chi}_1^0$ scenario is suppressed to
be less than 0.4  for the light-$h_1$ case. For the light-$a_1$ case,
most samples (about $96\%$) predict the same conclusion.
Since in the light-$h_1$ case the $\tilde\chi$-nucleon scattering
rate can reach the CoGeNT sensitivity,
this means that in the framework of the NMSSM the CoGeNT search for
the light dark matter will be correlated with the LHC
search for the Higgs boson via the di-photon channel. We checked that, once
the future XENON experiment fails in observing dark matter,
less than $1\%$ of the surviving samples in light $a_1$ case
predict the ratio of di-photon signal larger than 0.4.

\section{General extension for the explanation to PAMELA}\label{sec4}
To explain the PAMELA excess by dark matter annihilation, there
are some challenges. First,  dark matter must annihilate dominantly into leptons since PAMELA
has observed no excess of anti-protons  \cite{pamela} (However, as pointed
in Ref. \cite{kane}, this statement may be not so solid due to the significant
astrophysical uncertainties associated with their propagation).
Second, the explanation of PAMELA excess requires an annihilation rate which
is too large to explain the relic abundance if dark matter
is produced thermally in the early universe.
To tackle these difficulties, a new theory of dark matter was proposed in Ref. \cite{sommerfeld2}.
In this new theory the Sommerfeld effect of a new force in the dark sector can
greatly enhance the annihilation rate when the velocity of dark matter is much smaller
than the velocity at freeze-out in the early universe, and dark matter annihilates
into light particles which are kinematically allowed to decay to muons or electrons.

The above fancy idea is hard to realize in the MSSM, because there is not a
new force in the neutralino dark matter sector to induce the Sommerfeld enhancement
and neutralino dark matter annihilates largely to final
states consisting of heavy quarks or gauge and/or Higgs bosons~\cite{susy-dm-review,neu}.
However, as discussed in Ref. \cite{Hooper:2009gm}, in a general extension of the MSSM by
introducing a singlet Higgs superfield, the idea in Ref. \cite{sommerfeld2} can be
realized by the singlino-like neutralino dark matter:
\begin{itemize}
\item[(i)] The singlino dark matter annihilates to the light singlet Higgs bosons
and the relic density can be naturally obtained from the
interaction between singlino and singlet Higgs bosons.
\item[(ii)] The singlet Higgs bosons, not related to electro-weak symmetry breaking,
can be light enough to be kinematically allowed to
decay dominantly into muons or electrons through the tiny mixing with the Higgs doublets.
\item[(iii)] The Sommerfeld enhancement needed in  dark matter annihilation for the
explanation of PAMELA result can be induced by the light singlet Higgs boson.
\end{itemize}
In the following section, we will show how does this happen, the Higgs decay are also investigated.
\subsection{Higgs and neutralinos spectrum}
\label{gn1}
If introduce a singlet Higgs to the  MSSM in general, the renormalizable holomorphic
 superpotential of Higgs is given by Ref. \cite{Hooper:2009gm}
\bea
W = \mu \widehat{H}_u \cdot \widehat{H}_d+\l \widehat{S}
\widehat{H}_u \cdot \widehat{H}_d+\eta \widehat{S}
+\half \mu_s \widehat{S}^2 + \frac{1}{3} \k \widehat{S}^3 \ ,
\label{superpotential}
\eea
which include linear term, quadratic term, cubic term of singlet superfield
(like Wess-Zumino model \cite{wzmodel}). Note that in such case, we do not require
the singlet to solve the $\mu$ problem. The soft SUSY-breaking terms are given by
\bea
V_{\rm soft}&=& \tilde{m}_u^2 | H_u |^2 + \tilde{m}_d^2 | H_d |^2
    + \tilde{m}_s^2 | S |^2 \nn\\
&& +(B \mu H_u \cdot H_d +\l A_\l\ H_u \cdot H_d S
   + C\eta S +\half B_s \mu_s S^2 + \third \k A_\kappa\ S^3 + \mathrm{h.c.})\,.
\eea
After the Higgs fields develop the VEVs $v_u$, $v_d$ and $s$, i.e.,
we get the similar Higgs spectrum  as the NMSSM and the nMSSM which is
\begin{itemize}
\item[(1)] The CP-even Higgs mass matrix in the basis $(\phi_u, \phi_d, \sigma)$ is given by
\bea
{\cal M}_{h,11} & = & g^2 v_u^2 + \cot\beta\left[\l s (A_\l + \k s+\mu_s) +B\mu\right], \\
{\cal M}_{h,22} & = & g^2 v_d^2 + \tan\beta\left[\l s (A_\l + \k s+\mu_s) +B\mu\right], \\
{\cal M}_{h,33} & = & \l (A_\l+\mu_s) \frac{v_u v_d}{s}\, -\l \frac{\mu}{s} (v_u^2+v_d^2)
                      + \k s (A_\k + 4 \k s+3\mu_s)-\frac{C\eta}{s},\label{lth1}\\
{\cal M}_{h,12} & = & (2\l^2 - g^2) v_u v_d - \l s (A_\l + \k s +\mu_s)-B\mu, \\
{\cal M}_{h,13} & = & 2\l (\mu+\l s) v_u  - \l v_d (A_\l + 2\k s+\mu_s), \\
{\cal M}_{h,23} & = & 2\l (\mu+\l s) v_d - \l v_u (A_\l + 2\k s+\mu_s),
\eea
where $g^2 = (g_1^2 + g_2^2)/2$ with $g_1$ and $g_2$ being respectively the coupling constant
of SU(2) and U(1) in the SM.
\item[(2)] The CP-odd Higgs mass matrix ${\cal M}_a$ is given by
\bea
{\cal M}_{a,11} & = & (\tan\beta+\cot\beta)[\l s (A_\l + \k s+\mu_s)+B\mu], \\
{\cal M}_{a,22} & = & 4 \l \k v_u v_d + \l (A_\l+\mu_s) \frac{v_u v_d}{s}
             -\l\frac{\mu}{s}(v_u^2+v_d^2) \nn \\
             && -\k s(3A_\k+\mu_s)-\frac{C\eta}{s}-2B_s\mu_s, \label{lta1}\\
{\cal M}_{a,12} & = & \l \sqrt{v_u^2+v_d^2}\, (A_\l - 2\k s-\mu_s).
\eea
Note that here we have dropped the Goldstone mode, thus there
left a $2 \times 2$ mass matrix in the basis ($\tilde{A}, \xi$).
and it can be diagonalized by an orthogonal $2 \times 2$ matrix
$P'$ and the physical CP-odd states $a_i$ are given by
(ordered as $m_{a_1}<m_{a_2}$)
\bea
a_1 &=& P_{11}' \tilde{A} + P_{12}' S_I
    = P_{11}' (\cos\b \varphi_u + \sin\b \varphi_d ) +P_{12}'\xi ,  \\
a_2 &=& P_{21}' \tilde{A} + P_{22}' S_I
   = P_{21}' (\cos\b \varphi_u + \sin\b \varphi_d ) +P_{22}'\xi .
\eea
\item[(3)] The charged Higgs mass matrix ${\cal M}_\pm$ in the basis
$\left(H_u^+, H^+_d\right)$ is given by
\beq
{\cal M}_\pm = \left(\l s (A_\l + \k s+\mu_s) + B\mu + h_u h_d (\frac{g_2^2}{2} - \l^2)\right)
\left(\ba{cc} \cot\b & 1 \\ 1 & \tan\b \ea\right) ,
\eeq
\item[(4)] The neutralino mass matrix is :
\beq
{\cal M}_0 =
\left( \begin{array}{ccccc}
M_1          & 0             & m_Zs_W s_b    & - m_Z s_W c_b  & 0 \\
0            & M_2           & -m_Z c_W s_b  & m_Z c_W c_b    & 0 \\
m_Zs_W s_b   & -m_Z s_W s_b  & 0             & -\mu           & -\lambda v c_b \\
-m_Z s_W c_b & -m_Z c_W c_b  &  -\mu         & 0              & - \lambda v s_b \\
0            & 0             &-\lambda v c_b &- \lambda v s_b & 2 \k s+\mu_s
\end{array} \right). \label{neutralino_matrix}
\eeq
\end{itemize}

\subsection{ Explanation of PAMELA and implication on Higgs decays}
\label{gn2}
\begin{figure}[htbp]
\epsfig{file=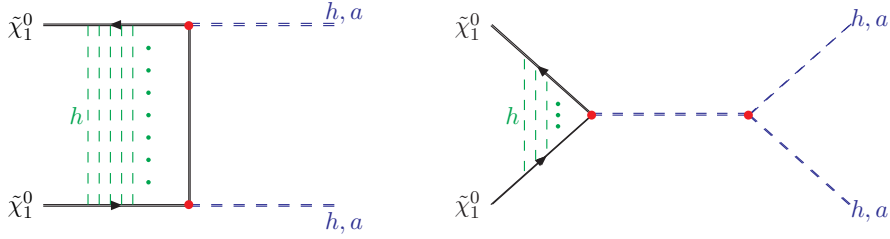,width=12cm}
 \vspace*{-0.5cm} \caption{Feynman
diagrams for singlino dark matter annihilation where Sommerfeld
enhancement is induced by exchanging $h$.
(taken from Ref. \cite{Wang:2009rj})} \label{fig9}
\end{figure}

To explain the observation of PAMELA, $a_1$ is singlet-dominant,
while $h_1$ is singlet-dominant
and the next-to-lightest $h_2$ is doublet-dominant ($h_{\rm SM}$).
 We use the notation:
\beq
a\equiv a_1, ~~~~h\equiv h_1, ~~~~h_{\rm SM}\equiv h_2.
\label{definition}
\eeq
As discussed in Ref. \cite{Hooper:2009gm}, when the lightest neutralino $\tilde\chi^0_1$
in Eq. (\ref{singlino}) is singlino-dominant, it can be a perfect candidate for
dark matter. As shown in Fig. \ref{fig9},
such singlino dark matter annihilates to a pair of light
singlet Higgs bosons followed by the decay $h\to aa$
($h$ has very small mixing with the Higgs doublets
and thus has very small couplings to the SM fermions).
In order to decay dominantly into muons,
$a$ must be light enough.
Further, in order to induce the Sommerfeld enhancement, $h$ must also be
light enough. From the superpotential term $\kappa \hat S^3$ we know
that the couplings $h\tilde\chi^0_1\tilde\chi^0_1$ and $a\tilde\chi^0_1\tilde\chi^0_1$
are proportional to $\kappa$. To obtain the relic density of dark matter, $\kappa$
should be ${\cal O}(1)$.
$h, a$ are singlet-dominant and $\tilde\chi^0_1$ is
singlino-dominant, this implies small mixing between singlet and
doublet Higgs fields. From the superpotential in
Eq.(\ref{superpotential}) we see that this means the mixing
parameter $\l$ must be small enough. On the other hand, from
Eq. (\ref{lth1}) and Eq. (\ref{lta1}) lightness of  $h_1$ and
$a_1$  also require $\l$ and other term approaching to zero.
Therefore, in our scan we require parameters $A_\k$ and $B_s$ has the
relation:
\bea
A_\k  &\sim & \left(-4 \k s-3\mu_s+\frac{C\eta}{\k s^2}\right),\\
2B_s \mu_s  &\sim & \left(-3A_\k \k s -\mu_s\k s-\frac{C\eta}{s}\right),
\eea
to realize light $h_1$ and $a_1$.

\begin{figure}[htbp]
\epsfig{file=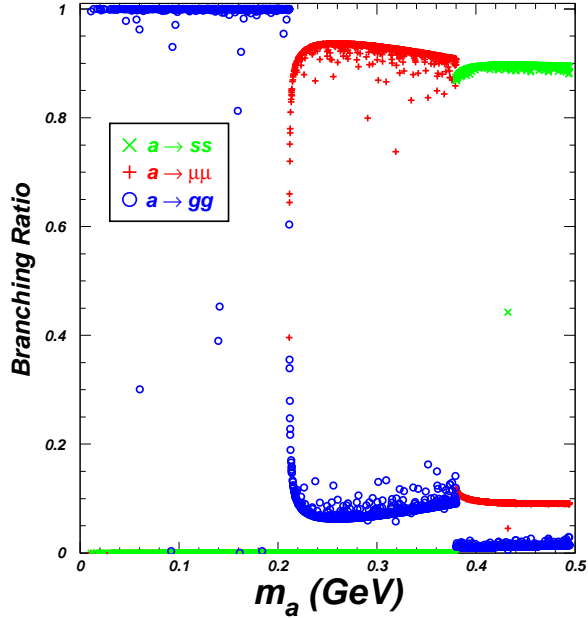,width=8cm}
\vspace*{-0.7cm}
\caption{The scatter plots showing the decay branching ratios
$a\to\mu^+\mu^-$ (muon), $a\to gg$ (gluon)
and $a\to s\bar s$ ($s$-quark) versus $m_a$
         for $\l=10^{-3}$. (taken from Ref. \cite{Wang:2009rj})}
\label{fig10}
\end{figure}
\begin{figure}[htbp]
\epsfig{file=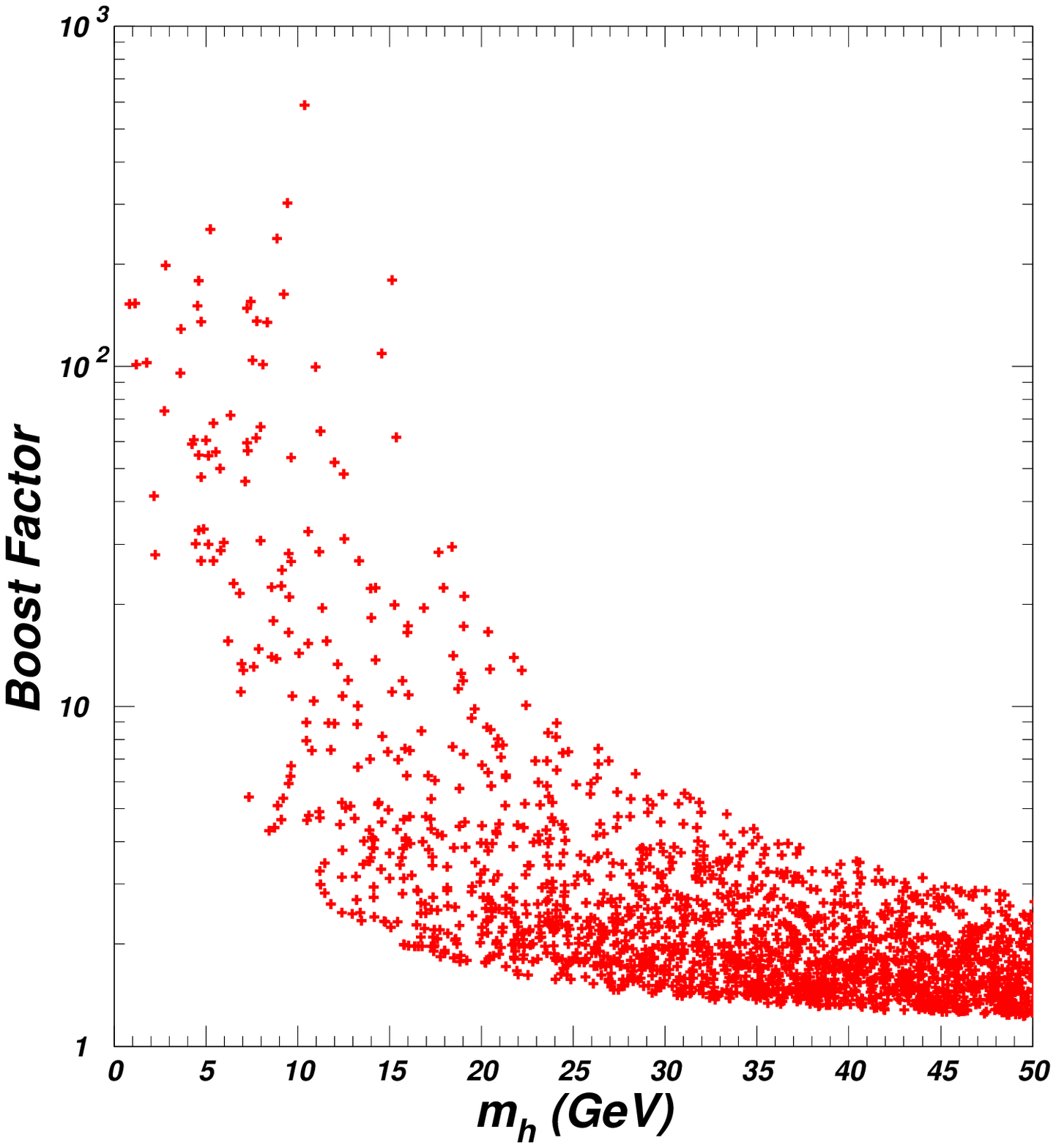,width=8cm} \vspace*{-0.7cm}
\caption{Same as Fig. \ref{fig10}, but
showing the Sommerfeld enhancement factor
 induced by $h$. (taken from Ref. \cite{Wang:2009rj})}
\label{fig11}
\end{figure}
\begin{figure}[htbp]
\epsfig{file=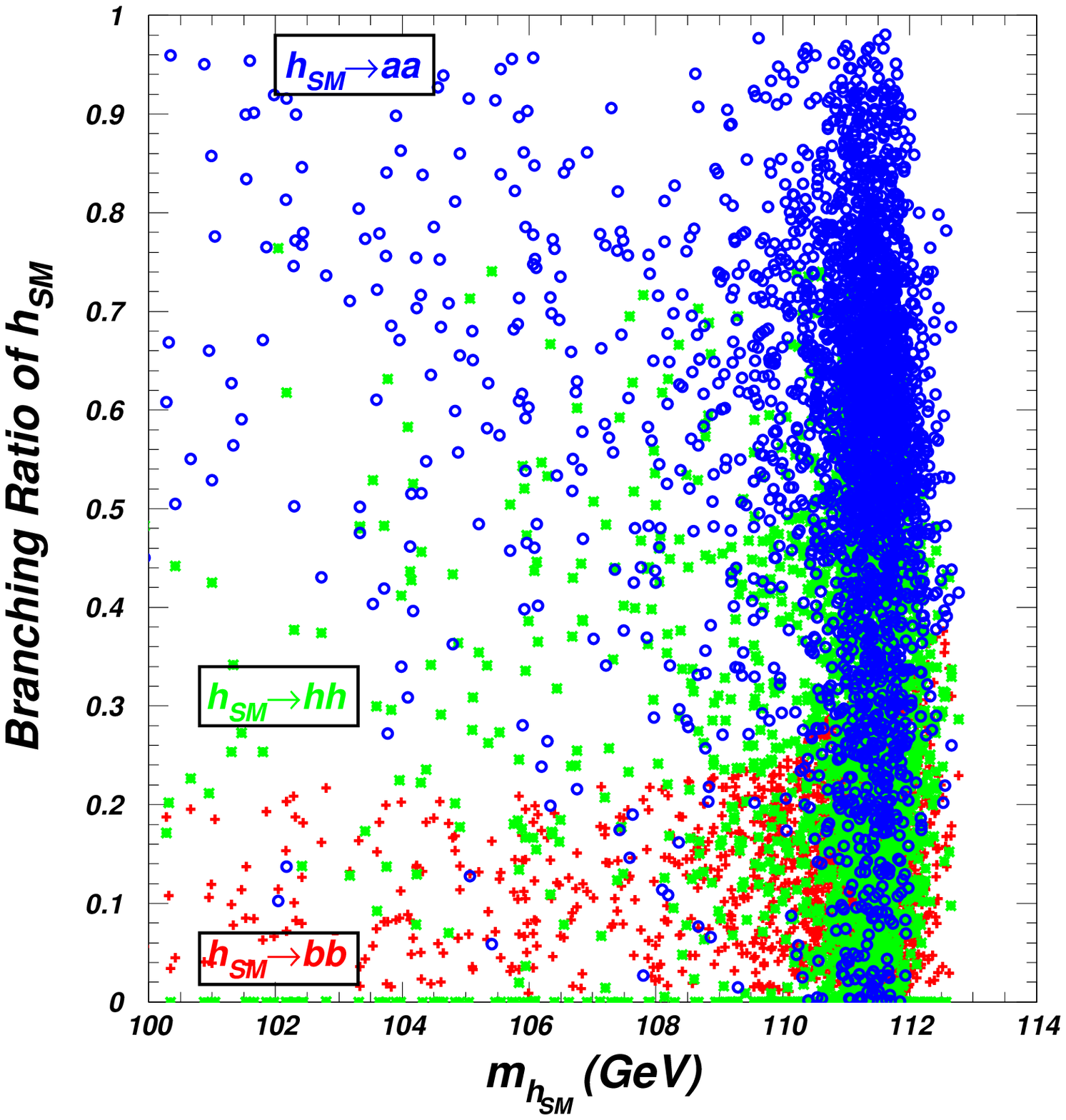,width=7cm}
\epsfig{file=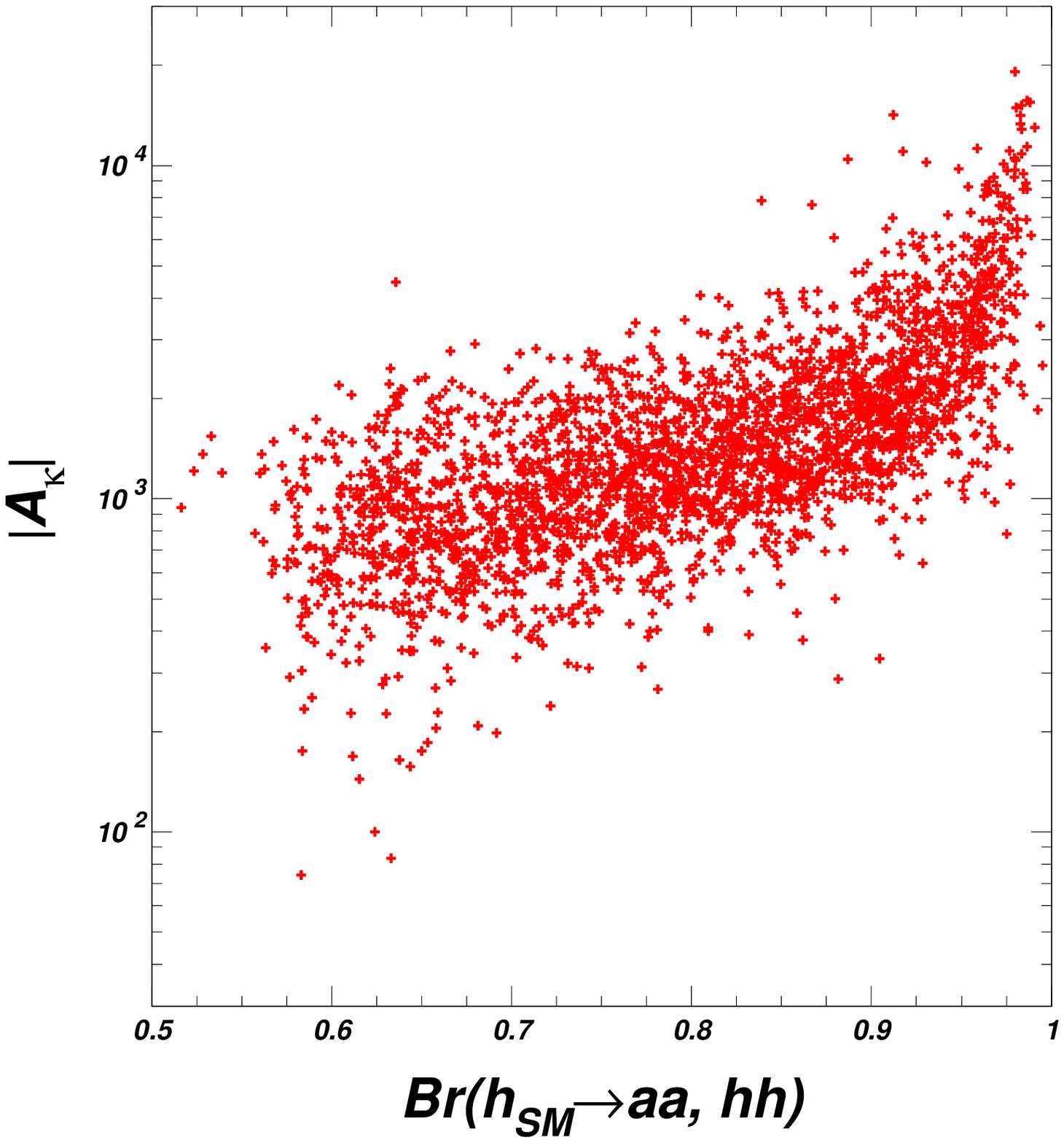,width=7cm}
\vspace*{-0.7cm}
\caption{Same as Fig. \ref{fig10}, but showing
branching ratio of $h_{\rm SM}\to a a, hh$ versus $m_{h_{\rm sm}}$
and $|A_\k|$ versus the branching ratio
         of $h_{\rm SM}\to a a, hh$. (taken from Ref. \cite{Wang:2009rj})}
\label{fig12}
\end{figure}

The numerical results of this model are displayed in different planes in
Figs.\ref{fig10}-\ref{fig12}. We see from Fig. \ref{fig10} that in the
range $2m_{\mu}<m_{a}<2m_{\pi}$, $a$ decays dominantly into muons.
It is clear  that  $h$ can be as light as a
few GeV, which is light enough to induce the necessary Sommerfeld
enhancement as shown in  Fig. \ref{fig11}.

In left plot of Fig. \ref{fig12}, we show the branching ratios of $h_{\rm SM}$ decays.
We see that in the allowed parameter space $h_{\rm SM}$ tends to decay into $aa$
or $hh$ instead of $b\bar b$. This can be understood as following,
the MSSM parameter space is stringently constrained by the LEP
experiments if $h_{\rm SM}$ is relatively light and decays dominantly
to $b\bar b$, and to escape such stringent constraints $h_{\rm SM}$ tends to have
exotic decays into  $aa$ or $hh$. As a result, the allowed parameter space tends
to favor a large $A_\kappa$, as shown in right plot of  Fig. \ref{fig12},
which greatly enhances the
couplings $h_{\rm SM}aa$ and $h_{\rm SM}hh$ through the soft term $\kappa A_\kappa S^3$
although $S$ has a small mixing with the doublet Higgs bosons.
Such an enhancement can be easily seen.
Take the coupling  $h_{\rm SM}hh$ as an example,
the soft term $\kappa A_\kappa S^3$ gives a term $\kappa A_\kappa \sigma^3$ which
then gives the interaction $\kappa A_\kappa ~(U_{13}^{H})^2 U^H_{23}~ h_{\rm SM}hh$
because $\sigma=U^H_{13}h_1+U^H_{23}h_2+U^H_{33}h_3$ with $h_1\equiv h$ and
$h_2\equiv h_{\rm SM}$ (see Eqs. (\ref{rotation}) and (\ref{definition})).
Although the mixing $(U_{13}^{H})^2U^H_{23}$ is small for
a small $\lambda$, a large $A_\kappa$ can enhance the coupling  $h_{\rm SM}hh$.
Note that as the mass of the observed Higgs boson at the LHC is around 125 GeV,
thus in the MSSM, the dominant decay mode of $h_{\rm SM}$ is $b\bar b$.
In this general singlet extension of the MSSM, its dominant decay
mode may be changed to $a a$ or $h h$, as shown in our above results.

Finally, we note that for the specified
singlet extensions like the nMSSM and the NMSSM, the
explanation of PAMELA and relic density through Sommerfeld
enhancement is not possible. The reason is that the parameter
space of such models is stringently constrained by various
experiments and dark matter relic density as shown in the above section, and,
as a result, the neutralino dark matter may explain either the
relic density or PAMELA, but impossible to explain both via
Sommerfeld enhancement. For example, in the nMSSM
various experiments and dark matter relic density constrain the
neutralino dark matter particle in a narrow mass range \cite{dm-nmssm},
 which is too light to explain PAMELA.
\section{Summary}\label{sec5}
At last we summarize here, the SUSY dark matter and Higgs physics
will be changed if introducing a singlet to the MSSM. Under the latest
results of  dark matter detection, we have:
\begin{enumerate}
\item In the MSSM, the NMSSM and the nMSSM,
the latest detection result can exclude a large
part of the parameter space allowed by current collider constraints
and the future SuperCDMS and XENON can cover most of the allowed
parameter space.
\item Under the new dark matter constraints, the singlet sector will
decouple from the MSSM-like sector in the NMSSM, thus the phenomenologies
of dark matter and Higgs are similar to the MSSM. The singlet sector
make the nMSSM quite different from the MSSM, the LSP in the
nMSSM are singlet dominant, and the SM-like Higgs will
 mainly decay into the singlet sector. Future precision measurements
will give us an opportunity to determine
whether the new scalar is from standard model or from SUSY.
Perhaps the nMSSM will be the first model be excluded
for its much larger branching ratio of invisible Higgs decay.
\item The NMSSM can allow light dark matter at several GeV exists.
Light CP-even or CP-odd Higgs boson must be present so as
to satisfy the measured dark matter relic density.
In case of the presence of a light CP-even Higgs boson,
the light neutralino dark matter can  explain the CoGeNT
and DAMA/LIBRA results.
Further, we find that in such a scenario the SM-like Higgs boson
will decay predominantly into a pair of light Higgs
bosons or a pair of neutralinos and the conventional decay
modes will be greatly suppressed.
\item The general singlet extension of the MSSM gives a perfect
explanation for both  the relic density and  the PAMELA result
through the Sommerfeld enhanced annihilation into singlet Higgs
bosons ($a$ or $h$ followed by $h\to a a$) with $a$ being light
enough to decay dominantly to muons or electrons.  Although the
light singlet Higgs bosons have small mixing with the Higgs
doublets in the allowed parameter space, their couplings with the
SM-like Higgs boson $h_{SM}$ can be enhanced by the soft parameter $A_\kappa$.
In order to meet the stringent LEP constraints, the $h_{SM}$ tends to decay into
the singlet Higgs pairs $aa$ or $hh$ instead of $b\bar b$.
\end{enumerate}
\section*{Acknowledgment}
This work was supported in part by the NSFC No. 11005006, No. 11172008 and Doctor
Foundation of BJUT No. X0006015201102.

\end{document}